\documentstyle[12pt,epsf,fleqn]{article}
\newcommand{\fips}[1]{\epsffile{#1.eps}}
\newcommand{\be}{\begin{equation}}
\newcommand{\ee}{\end{equation}}
\newcommand{\lee}[1]{\label{#1} \end{equation}}
\newcommand{\bea}{\begin{eqnarray}}
\newcommand{\leea}[1]{\label{#1} \end{eqnarray}}
\newcommand{\eea}{\end{eqnarray}}
\newcommand{\nn}{\nonumber}
\newcommand{\ga}{\gamma}
\newcommand{\de}{\delta}

\newcommand{\la}{\lambda}

\newcommand{\Ga}{\Gamma}
\newcommand{\De}{\Delta}
\newcommand{\La}{\Lambda}

\begin{document}


\title{Modified similarity renormalization of Hamiltonians.\\
QED on the light front. \vspace*{.5cm}}
\author{E.~L.~Gubankova\thanks{Posters presented on the workshop
"Confinement, duality and non-perturbative aspects of QCD",
23 June-4 July,Cambridge,England},
\vspace*{.5cm}\\
\normalsize\it Institut f\"ur Theoretische Physik der Universit\"at Heidelberg \\
\normalsize\it Philosophenweg 19, D69120 Heidelberg, FRG}

\maketitle

\vspace*{1cm}
\begin{abstract}
Modified similarity renormalization of Hamiltonians is proposed,
that performes by means of flow equations the similarity
transformation of Hamiltonian in the particle number space.
This enables to renormalize in the energy space
the field theoretical Hamiltonian and makes possible to work
in a severe trancated Fock space for the renormalized Hamiltonian.
\end{abstract}

\newpage
\section{Motivation}

How to solve relativistic bound state problem numerically?
\be
H|\psi>=E|\psi>
\ee
H-canonical Hamiltonian of the field theory.

{\bf Problem}\\

\underline{H-is infinite dimensional}   
\\
\parbox{8.5cm}{\sloppy 1.in 'energy' space

(high-energy fluctuations (states) couple to low-energy one)}
\hfill
\parbox{8.5cm}{\sloppy 2.in 'particle number'\\
space (there are\\ interactions in Hamiltonian,
which change particle number of a state; 
therefore in Fock space representation
many-body states couple to few-body states)}
\\
{\bf The methods available} to handle\\

\parbox{8.5cm}{\sloppy \underline{the first problem:}

1.Regularization and similarity renormalization of Hamiltonians
by Glazek,Wilson

2.Projection of high-energy states onto the low-energy one
(Bloch-Fleshbach formalism)}
\hfill
\parbox{8.5cm}{\sloppy \underline{the second problem:}

1.Tamm-Dancoff trancation

2.Method of iterrative resolvents by Pauli}

The field theoretical Hamiltonian has the 'pentadiagonal' form
in the 'particle number' representation, each 'particle number' block 
in its turn has infinite 
many states in 'energy' representation.
\\
{\bf The way out} to construct effective \underline{finite dimensional},
in 'energy' and 'particle number' space, Hamiltonian,which controls 
the effects of high-energy and many-body states.

\begin{figure}
$$
\fips{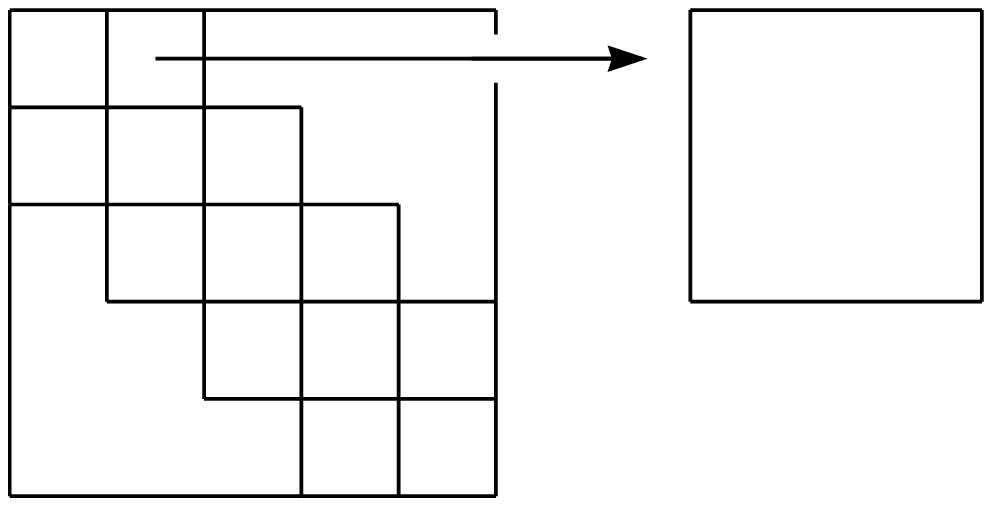}
\setlength{\unitlength}{0.240900pt}
\begin{picture}(0,0)
\put(-1000,650){\makebox(0,0){particle number}}
\put(-250,650){\makebox(0,0){energy}}
\end{picture}
$$
\begin{center}
Figure 1: pentadiagonal form of Hamiltonian
\end{center}
\label{Hamiltonian}
\end{figure}

\newpage
\section{Similarity renormalization in 'energy' and 
'particle number' space (compared)}

{\bf I.Idea}
\\
\parbox{8.5cm}{\sloppy \underline{'Energy renormalization'}}
\hfill
\parbox{8.5cm}{\sloppy \underline{'Particle number renormalization'}}
\\
{\bf Program} (Glazek,Wilson)
\\
\underline{1.With the help of similarity transformation decouple}
\\
\parbox{8.5cm}{\sloppy high-energy\\ and low-energy states}
\hfill
\parbox{8.5cm}{\sloppy many-body\\ and few-body states}
\\
\underline{2.Simulate the effects of}
\\
\parbox{8.5cm}{\sloppy high-energy states}
\hfill
\parbox{8.5cm}{\sloppy many body states}
\\
by a set of local operators-counterterms; and new induced interactions,
corresponding to marginal relevant operators of the theory
\\
{\bf Aim}
\\
\underline{Effective Hamiltonian $H^{eff}$ is}
\\
\parbox{8.5cm}{\sloppy band-diagonal in 'energy' space\\
$|E_i-E_j|<\la$}
\hfill
\parbox{8.5cm}{\sloppy block-diagonal in 'particle number' space\\
each block conserves the number of particles}
\\
{\bf Choice}
\\
\underline{1.'Diagonal' part of Hamiltonian $H_d$ is}
\\
\parbox{8.5cm}{\sloppy $H_0$-free (noninteracting) part of Hamiltonian}
\hfill
\parbox{8.5cm}{\sloppy $H_c$-particle number conserving part 
of Hamiltonian}
\\
\underline{2.Basis}
\\
\parbox{8.5cm}{\sloppy $H_0|i>=E_i|i>$\\
$|i>$-single particle state}
\hfill
\parbox{8.5cm}{\sloppy $H_c|k>=E_k|k>$\\
$|k>$-state with particle number 'k'}
\\
\underline{Similarity function in 'diagonal' sector is equal to unity}
$u_d =1$

{\bf Tool}
\\
\underline{Flow equations of Wegner}
\\
\parbox{8.5cm}{\sloppy in 'energy space'}
\hfill
\parbox{8.5cm}{\sloppy in 'particle number space'}

\begin{figure}
$$
\fips{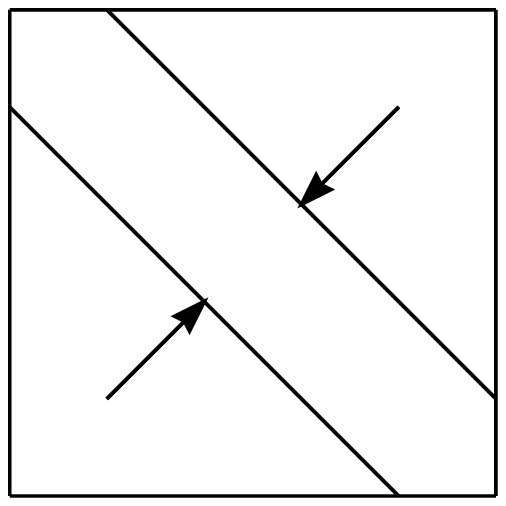}
\setlength{\unitlength}{0.240900pt}
\begin{picture}(0,0)
\put(-200,480){\makebox(0,0){$\lambda$}}
\end{picture}
$$
\begin{center}
Figure 2: band-diagonal Hamiltonian in 'energy space'
\end{center}
\label{Hamiltonian}
\end{figure}
\begin{figure}
$$
\fips{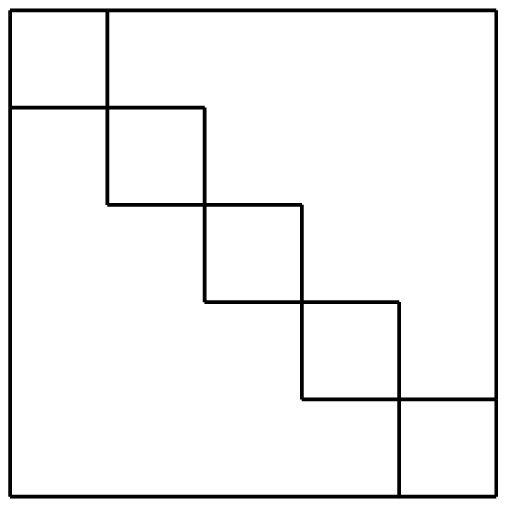}
\setlength{\unitlength}{0.240900pt}
\begin{picture}(0,0)
\end{picture}
$$
\begin{center}
Figure 3: block-diagonal Hamiltonian in 'particle number space'
\end{center}
\label{Hamiltonian}
\end{figure}

\newpage
{\bf II.Technique}
\\
\bea
&& H_0,H_c\rightarrow H_d\nn\\
&& |i>,|k>\rightarrow |n>\nn\\
&& H=H_d+H_r
\eea
\\\
Flow equation (Wegner)
\bea
&& \frac{dH(l)}{dl}=[\eta(l),H(l)]\nn\\
&& \eta=[H_d,H_r]
\eea
In matrix form
\bea
&& \frac{dH_{mn}}{dl}=[\eta,H_r]_{mn}-(E_m-E_n)^2H_{rmn}\nn\\
&& \eta_{mn}=(E_m-E_n)H_{rmn}
\eea
Through the similarity function
\bea
&& \frac{dH_{mn}}{dl}=[\eta,H_r]_{mn}+
\frac{du_{mn}}{dl}\frac{H_{mn}}{u_{mn}}\nn\\
&& \eta_{mn}=\frac{1}{E_m-E_n}(-\frac{du_{mn}}{dl}
\frac{H_{mn}}{u_{mn}})
\eea
where the similarity function is
\bea
&& u_{mn}(l)=\exp(-\De_{mn}^2l)\nn\\
&& \De_{mn}=E_m-E_n
\eea

{\bf Perturbation theory} $H^{(n)}\sim e^n$\\
{\bf 'rest sector'} (matrix elements with $m\neq n$)
\bea
&& \frac{dH_r^{(1)}}{dl}=-(E_m-E_n)^2H_r^{(1)}\nn\\
&& H_r^{(1)}(l)=u_{mn}(l)H_r^{(1)}(0)\nn\\
&& \eta_{mn}^{(1)}=(E_m-E_n)H_r^{(1)}
\eea
\bea
&& \frac{dH_r^{(2)}}{dl}=[\eta^{(1)},H_r^{(1)}]_r
-(E_m-E_n)^2H_r^{(2)}\nn\\
&& H_r^{(2)}(l)=u_{mn}(l)\tilde{H}_r^{(2)}(l)\nn\\
&& \tilde{H}_r^{(2)}(l)=\tilde{H}_r^{(2)}(0)+
\int_0^l\frac{1}{u_{mn}}[\eta^{(1)},H_r^{(1)}]_rdl'
\eea
{\bf 'diagonal sector'} (matrix elements with $m=n$)
\bea
&& \frac{dH_d^{(2)}}{dl}=[\eta^{(1)},H_r^{(1)}]_d\nn\\
&& H_d^{(2)}(l)=H_d^{(2)}(0)+\int_0^l[\eta^{(1)},H_r^{(1)}]_d dl'
\eea
1.\\
\parbox{8.5cm}{\sloppy {\bf 'energy renormalization'}\\
(Glazek,Wilson similarity renormalization)\\
$l=1/\la^2$ ($\la$-UV-cutoff),\\
$\forall$ matrix elements
holds $|E_i-E_j|<\la$\\
\underline{$H^{eff}$ is band-diagonal}}
\hfill
\parbox{8.5cm}{\sloppy {\bf 'particle number renormalization'}\\
when $l\rightarrow\infty$ ($\la\rightarrow 0$)
the 'rest sector' is completly eliminated\\
\underline{$H^{eff}$ is block-diagonal}}
\\
2.in 'diagonal sector' similarity function is equal to unity
$u_d=1$\\
3.in the case of QED $H^{(2)}$ simulate (to the leading order)
the effects of high-energy
(many-body) states.

\newpage
\section{Modified similarity renormalization.}

{\bf I.Idea}
\\
{\bf Aim}
\\
Perform 'energy' and 'particle number' renormalization simultaneously.
Simulate both effects of high-energy and many-body states.
\\
{\bf Program}
\\
Perform similarity transformation in the 'particle number space'
to bring Hamiltonian to a block-diagonal form, with the number of
particles conserving in each block.
\\
{\bf Tool}
\\
Flow equations in the 'energy space', organized in a way to eliminate 
the particle number changing sectors and to generate effective Hamiltonian
in the particle number conserving sectors.
The effects of high-energy and many-body states are simulated then
by a set of effective interactions, which do not change particle number
and do not couple to high-energy states either.
\\
{\bf Coice}
\\
1.\underline{$H_d$ (diagonal part)} 
is the particle number conserving part of the Hamiltonian\\
that is equivalent to\\ 
\underline{Similarity function} in 
'diagonal particle number sector' is equal to unity
\be
H_d=H_c~~\Leftrightarrow~~u_{dij}=1
\ee

\underline{2.Basis}
\be
H_0|i>=E_i|i>
\ee
$H_0$-free (noninteracting) part of Hamiltonian,

$|i>$-single particle state
\\
'Energy' and 'particle number' renormalizations are closely related 
in a comlicated way.

\begin{figure}
$$
\fips{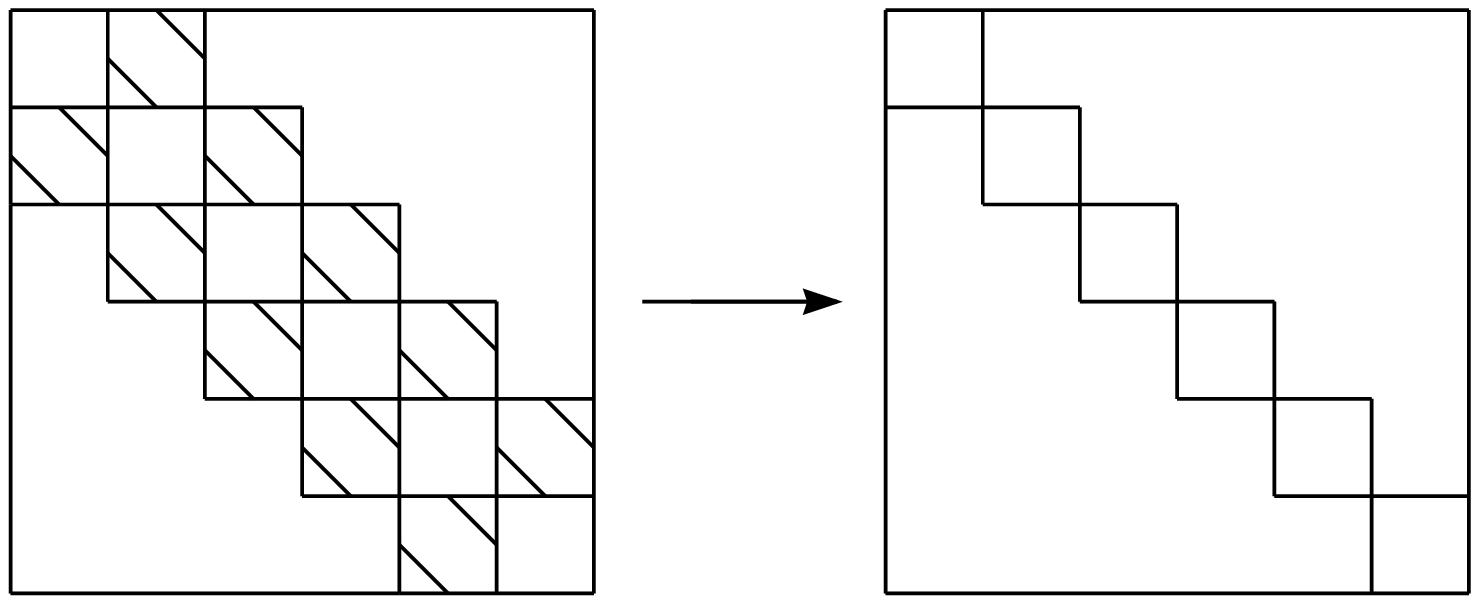}
\setlength{\unitlength}{0.240900pt}
\begin{picture}(0,0)
\put(-1500,780){\makebox(0,0){particle number}}
\put(-900,420){\makebox(0,0){U}}
\end{picture}
$$
\begin{center}
Figure 4: Modified similarity renormalization of Hamiltonians
\end{center}
\label{Hamiltonian}
\end{figure}

\newpage

{II.\bf Technique}
\bea
&& H=H_d+H_r\nn\\
&& \eta=[H_d,H_r]
\eea
\\
$H_d$-particle number conserving part ('diagonal sector')\\
$H_r$-particle number changing part ('rest sector')
\\
Perturbation theory
\bea
&& H=H_{0d}+\sum_n(H_d^{(n)}+H_r^{(n)})\nn\\
&& H^{(n)}\sim e^n
\eea
\bea
&& \frac{dH^{(n)}}{dl}=\sum_k[\eta^{(k)},H_d^{(n-k)}+H_r^{(n-k)}]\nn\\
&& +\sum_k[[H_d^{(k)},H_r^{(n-k)}]H_{0d}]+[[H_{0d},H_r^{(n)}]H_{0d}]\nn\\
&& \eta^{(n)}=[H_{0d},H_r^{(n)}]+\sum_k[H_d^{(k)},H_r^{(n-k)}]
\eea
In matrix form
\bea
&& H_{0d}|i>=E_i|i>\nn\\
&& \frac{dH_{ij}^{(n)}}{dl}=\sum_k[\eta^{(k)},H_d^{(n-k)}+
H_r^{(n-k)}]_{ij}\nn\\
&& -(E_i-E_j)\sum_k[H_d^{(k)},H_r^{(n-k)}]_{ij}-(E_i-E_j)^2H_{rij}^{(n)}
\eea
$|i>$-single particle state
\\
\underline{'diagonal sector'}
\bea
&& \frac{dH_{dij}^{(n)}}{dl}=\sum_k[\eta^{(k)},H_d^{(n-k)}+
H_r^{(n-k)}]_{dij}\nn\\
&& +\sum_k[[H_d^{(k)},H_r^{(n-k)}]_d,H_{0d}]_{ij}
\eea
\underline{'rest sector'}
\bea
&& \frac{dH_{rij}^{(n)}}{dl}=\sum_k[\eta^{(k)},H_d^{(n-k)}+
H_r^{(n-k)}]_{rij}\nn\\
&& +\sum_k[[H_d^{(k)},H_r^{(n-k)}]_r,H_{0d}]_{ij}
-(E_i-E_j)^2H_{rij}^{(n)}
\eea
\\
{\bf 'diagonal sector'}
new terms are induced;
\\
in 'diagonal sector' the matrix elements with 
any energy differences are present
\\
{\bf 'rest sector'}
\\
has band-diagonal structure in the 'energy space'
\bea
&& H_{rij}=u_{ij}\tilde{H}_{rij}\nn\\
&& u_{ij}=\exp(-(E_i-E_j)^2l)
\eea
when $l\rightarrow\infty$ the 'rest part' is completly eliminated
(exept the diagonal in 'energy space' matrix elements $i=j$, which
do not contribute to physical values)
\\
the effective Hamiltonian has block-diagonal form in the 'particle number'
space.
  
\newpage
\section{Bound states in light-front QED (LFQED).}

{\bf I.Motivation.}
\\
QED-perturbation theory in bare coupling constant is applicable;\\
LFQED-justify the 'parton picture' of bound states as a weakly bound
system of the constituents.
{\bf II.Positronium on the light-front.}
\\
The complete elimination (flow equation in the 'energy space'
runs up to the limit value of\\
$l\rightarrow\infty$ )($\la\rightarrow 0)$)
of the electron-photon vertex ($|ee\ga>$ sector) gives rise
to the new generated electron-positron interaction ($|e\bar{e}>$ sector).
The instantaneous interaction (artefact of light-front formulation)
stays intact to the leading order by this flow as the particle
number conserving interaction.

\underline{In exchange channel}\\
$p_1+p_2\rightarrow p_3+p_4$,\\ 
$p_1=(x,k_{\perp}),p_3=(x',k'_{\perp})$-electron
\bea
&& V^{gen}=-e^2M_{2ii}^{ex}\frac{1}{(p_1^+-p_3^+)}
\left( \frac{\int_0^{\infty} df_{p_1p_3\la'}/d\la'
 f_{p_4p_2\la'}d\la'}
 {\De_{p_1p_3}} \right. \nn\\
&&\hspace{8cm} + \left. \frac{\int_0^{\infty}
 df_{p_4p_2\la'}/d\la'f_{p_1p_3\la'}d\la'}{\De_{p_4p_2}}
\right) \nn\\
&& V^{inst}=-\frac{4e^2}{(p_1^+-p_3^+)^2}\de_{s_1s_3}\de_{s_2s_4}
\eea
where $M_{2ii}^{ex}$ is the matrix element of two $ee\ga$ verteces
in exchange channel, $M_{2ii}^{ex}=\Ga(p_1,p_3)\Ga(-p_4,-p_2)$;\\
$\De_{p_1p_3},\De_{p_4p_2}$ are energy denominators,
$\De_{p_1p_3}=p_1^--p_3^--(p_1-p_3)^-$;

$f_{p_1p_3},f_{p_4p_2}$ define the velocities of elimination of
corresponding $ee\ga$ verteces in the 'energy space'
\be
e_{p_1p_3}(\la)=e(\La)\frac{f_{p_1p_3\la}}{f_{p_1p_3\La}}
\ee
$e(\La)$ is the bare coupling constant;
$f_{p_1p_3\la}$ depends on the transformation performed and is the function
of the 'similarity function' $u_{p_1p_3\la}$;
in the simplest case $f_{p_1p_3\la}=u_{p_1p_3\la}$ 
(ansatz of 

Wegner $u_{p_1p_3}=\exp(-\frac{\De_{p_1p_3}^2}{\la^2})$,

Glazek,Wilson $u_{p_1p_3}=\Theta(\la-|\De_{p_1p_3}|)$).
\\
The electron-positron interaction is (neglecting annihilation channel)
\be
V=V^{gen}+V^{inst}
\ee

{\bf Properties of 'V'}
\\
1.Generated interaction together with the instantaneous one insure
the attractive interaction in the whole parameter range of momenta.
To the leading order of nonrelativistic approximation 'V' containes
the Coulomb term and spin-dependent interaction, which
gives rise to the correct value of singlet-triplet splitting
$\frac{7}{6}\alpha^2Ryd$.
\\
2.'V' is finite in the collinear limit $|x-x'|\rightarrow 0$.
\\
3.'State-independence' of counterterms (preliminary).
By the numerical calculation the same counterterms must be introduced
for the ground and first exited (in quantum number 'n') states
to get the cutoff independent physical masses.

\newpage
\section{Advantages of MSR.}
{\bf I.MSR compared to Tamm-Dancoff trancation (TD)
and method of iterrative resolvents (IR).}
\\
1.Many-body and few-body states are decoupled in MSR. This enables
to work in a restricted Fock (particle number) space and not to encounter
the usual difficulties of TD and IR. Namely, the counterterms to be
introduced are 'sector-independent' (as compared to TD)
and 'state-independent' (as compared to IR).
\\
2.The 'energy renormalization' is performed simultaneously
with the 'particle number renormalization'. Therefore all counterterms
to the definite order in bare coupling, associated with canonical
operators of the theory (relevant and marginal) and also
possibly new induced marginal relevant operators,
are obtained automatically in the procedure. No additional calculation
of perturbation theory corrections is needed.
\\
{\bf II.MSR compared to similarity renormalization (SR) 
of Glazek,Wilson.}
\\
1.Restriction of Fock space is no more dangerous.
The 'state-independent' counterterms are to be introduced.
\\
2.The artificial cutoff $\la$ (the size of the band $|E_i-E_j|<\la$)
is not needed.The procedure of MSR is performed to the 'end'
(the limit $\la\rightarrow 0$ ($l\rightarrow \infty$) 
of  'MSR' effective Hamiltonian is well defined),
where the 'particle number' changing interactions are completly
eliminated.Therefore,

first,no cutoff dependence (except the renormalization group 
running of couplings) is present in the effective (renormalized) 
Hamiltonian and hence the physical values;

second,there is no ambiguity in the choice of the step,
where the procedure must be stoped (namely in the choice of the 
final value of $\la$);

third,no difficulties on the convergency of the perturbation theory occure.
Namely,the effective Hamiltonian containes the correct value
of the zero'th approximation to calculate the bound states.
\\
3.The elimination of the new induced interactions, corresponding
to the marginal relevant operators of the theory, causes the 
convergency problem of SR. This was observed by Wegner on
the example of $1$-dimensional problem using flow equations.
The convergency of the procedure in MSR is well defined.

\end{document}